\documentclass[tight,twocolumn]{aastex62}
\usepackage{apjfonts}
\usepackage{psfig}
\usepackage{float}

\usepackage{graphics}
\usepackage{amssymb,amsmath}
\usepackage{color}
\usepackage{todonotes}
\usepackage{xspace}
\usepackage{comment}
\newcommand{\name}{ASASSN-18ey\xspace}

\newcommand{\degree}{\ensuremath{^\circ}}

\newcommand{\swift}{\textit{Swift}\xspace}
 
\newcommand{\IRAF}{\textsc{iraf}}

\begin{document}

\title{ASASSN-18ey: The Rise of a New Black-Hole X-ray Binary}  
\shorttitle{ASASSN-18ey} 
\shortauthors{Tucker, Shappee, Holoien et al. }

\author[0000-0002-2471-8442]{M. A. Tucker}
\affiliation{Institute for Astronomy, University of Hawai'i, 2680 Woodlawn Drive, Honolulu, HI 96822, USA}

\author[0000-0003-4631-1149]{B.~J.~Shappee}
\affiliation{Institute for Astronomy, University of Hawai'i, 2680 Woodlawn Drive, Honolulu, HI 96822, USA}

\author[0000-0001-9206-3460]{T.~W.-S.~Holoien}
\altaffiliation{Carnegie Fellow}
\affiliation{The Observatories of the Carnegie Institution for Science, 813 Santa Barbara St., Pasadena, CA 91101, USA}

\author{K.~Auchettl}
\affiliation{Center for Cosmology and AstroParticle Physics (CCAPP), The Ohio State University, 191 W.\ Woodruff Ave., Columbus, OH 43210, USA}
\affiliation{Department of Physics, The Ohio State University, 191 W. Woodruff Avenue, Columbus, OH 43210, USA}
\affiliation{Dark Cosmology Centre, Niels Bohr Institute, University of Copenhagen, Blegdamsvej 17, 2100 Copenhagen, Denmark}

\author{J. Strader}
\affiliation{Department of Physics and Astronomy, Michigan State University, 567 Wilson Rd, East Lansing, MI 48824, USA}

\author{K.~Z.~Stanek}
\affiliation{Center for Cosmology and AstroParticle Physics (CCAPP), The Ohio State University, 191 W.\ Woodruff Ave., Columbus, OH 43210, USA}
\affiliation{Department of Astronomy, The Ohio State University, 140 West 18th Avenue, Columbus, OH 43210, USA}

\author{C.~S.~Kochanek}
\affiliation{Center for Cosmology and AstroParticle Physics (CCAPP), The Ohio State University, 191 W.\ Woodruff Ave., Columbus, OH 43210, USA}
\affiliation{Department of Astronomy, The Ohio State University, 140 West 18th Avenue, Columbus, OH 43210, USA}

\author{A. Bahramian}
\affiliation{Department of Physics and Astronomy, Michigan State University, 567 Wilson Rd, East Lansing, MI 48824, USA}
\affiliation{International Centre for Radio Astronomy Research Curtin University, GPO Box U1987, Perth, WA 6845, Australia}

\collaboration{}
\collaboration{\textit{ASAS-SN}}


\author{Subo~Dong}
\affiliation{Kavli Institute for Astronomy and Astrophysics, Peking University, Yi He Yuan Road 5, Hai Dian District, Beijing 100871, China}

\author[0000-0003-1072-2712]{J.~L.~Prieto}
\affiliation{N\'ucleo de Astronom\'ia de la Facultad de Ingenier\'ia y Ciencias, Universidad Diego Portales, Av. Ej\'ercito 441, Santiago, Chile}
\affiliation{Millennium Institute of Astrophysics, Santiago, Chile}
\author{J.~Shields}
\affiliation{Department of Astronomy, The Ohio State University, 140 West 18th Avenue, Columbus, OH 43210, USA}

\author{Todd A.~Thompson}
\affiliation{Center for Cosmology and AstroParticle Physics (CCAPP), The Ohio State University, 191 W.\ Woodruff Ave., Columbus, OH 43210, USA}
\affiliation{Department of Astronomy, The Ohio State University, 140 West 18th Avenue, Columbus, OH 43210, USA}

\author[0000-0002-0005-2631]{John F. Beacom}
\affiliation{Center for Cosmology and AstroParticle Physics (CCAPP), The Ohio State University, 191 W.\ Woodruff Ave., Columbus, OH 43210, USA}
\affiliation{Department of Physics, The Ohio State University, 191 W. Woodruff Avenue, Columbus, OH 43210, USA}
\affiliation{Department of Astronomy, The Ohio State University, 140 West 18th Avenue, Columbus, OH 43210, USA}

\author{L.~Chomiuk}
\affiliation{Department of Physics and Astronomy, Michigan State University, 567 Wilson Rd, East Lansing, MI 48824, USA}

\collaboration{}
\collaboration{\textit{ATLAS}}

\author{L.~Denneau}
\affiliation{Institute for Astronomy, University of Hawai'i, 2680 Woodlawn Drive, Honolulu, HI 96822, USA}

\author{H.~Flewelling}
\affiliation{Institute for Astronomy, University of Hawai'i, 2680 Woodlawn Drive, Honolulu, HI 96822, USA}

\author{A.~N.~Heinze}
\affiliation{Institute for Astronomy, University of Hawai'i, 2680 Woodlawn Drive, Honolulu, HI 96822, USA}

\author{K.~W.~Smith}
\affiliation{Astrophysics Research Centre, School of Mathematics and Physics, Queens University Belfast, Belfast BT7 1NN, UK.}

\author{B.~Stalder}
\affiliation{LSST, 950 North Cherry Avenue, Tucson, AZ 85719, USA}

\author{J.~L.~Tonry}
\affiliation{Institute for Astronomy, University of Hawai'i, 2680 Woodlawn Drive, Honolulu, HI 96822, USA}

\author{H.~Weiland}
\affiliation{Institute for Astronomy, University of Hawai'i, 2680 Woodlawn Drive, Honolulu, HI 96822, USA}

\author{A. Rest}
\affiliation{Space Telescope Science Institute, 3700 San Martin Drive,
Baltimore, MD 21218, USA}

\collaboration{}

\author{M. E. Huber}
\affiliation{Institute for Astronomy, University of Hawai'i, 2680 Woodlawn Drive, Honolulu, HI 96822, USA}

\author[0000-0003-2431-981X]{D. M. Rowan}
\affiliation{Institute for Astronomy, University of Hawai'i, 2680 Woodlawn Drive, Honolulu, HI 96822, USA}
\affiliation{Department of Physics and Astronomy, Haverford College, 370 Lancaster Avenue, Haverford, Pennsylvania 19041, USA}

\author{K. Dage}
\affiliation{Department of Physics and Astronomy, Michigan State University, 567 Wilson Rd, East Lansing, MI 48824, USA}

\correspondingauthor{Michael A. Tucker}
\email{tuckerma@hawaii.edu}

\date{Accepted XXX. Received YYY; in original form ZZZ}



\begin{abstract}
We present the discovery of ASASSN-18ey (MAXI J1820+070), a new black hole low-mass X-ray binary discovered by the All-Sky Automated Survey for SuperNovae (ASAS-SN). A week after ASAS-SN discovered ASASSN-18ey as an optical transient, it was detected as an X-ray transient by MAXI/GCS.  Here, we analyze ASAS-SN and Asteroid Terrestrial-impact Last Alert System (ATLAS) pre-outburst optical light curves, finding evidence of intrinsic variability for several years prior to the outburst. While there was no long-term rise leading to outburst, as has been seen in several other systems, the start of the outburst in the optical preceded that in the X-rays by $7.20\pm0.97~\rm days$. We analyze the spectroscopic evolution of ASASSN-18ey from pre-maximum to $> 100~\rm days$ post-maximum. The spectra of ASASSN-18ey exhibit broad, asymmetric, double-peaked H$\alpha$ emission. The Bowen blend ($\lambda \approx 4650$\AA) in the post-maximum spectra shows highly variable double-peaked profiles, likely arising from irradiation of the companion by the accretion disk, typical of low-mass X-ray binaries. The optical and X-ray luminosities of ASASSN-18ey are consistent with black hole low-mass X-ray binaries, both in outburst and quiescence.
\end{abstract}



\vspace{1cm}
\section{Introduction}\label{sec:intro}

Low-mass X-ray binaries (LMXBs) consist of compact objects, either a neutron star (NS) or a black hole (BH), accreting material from a donor star with a typical mass of $M_{\rm{donor}} \lesssim 1~M_\odot$. The compact object is surrounded by an accretion disk fed by a donor star undergoing Roche Lobe overflow (RLOF). Observationally, LMXBs can be classified as either transient/outbursting sources or persistent/non-outbursting sources, with the caveat that transient LMXBs can go undetected for years or decades while in quiescence since their X-ray luminosity is low ($L_X\sim 10^{32}~\rm{erg}\;\rm s^{-1})$. During an outburst, these systems increase by several orders of magnitude in both X-ray and optical luminosity, routinely leading to their discovery. Conversely, persistent X-ray binaries have higher continuous X-ray luminosities ($L_X\sim 10^{36-38}~\rm{erg}\;\rm s^{-1}$), and the majority of these sources have been discovered by all-sky X-ray surveys. 
 
Transient LMXBs can be grouped into one of three categories based on their X-ray spectral state: high/soft/thermal, low/hard and very high/steep power law \citep[see ][for an overview of BH LMXB X-ray properties]{remillard06}. The high/soft state is dominated by thermal disk emission, with little to no power law component. Conversely, the low/hard state is dominated by the power law emission, contributing $\gtrsim 80\%$ of the observed flux. Finally, the very high state is characterized by a steep power law ($\Gamma > 2$) and usually dominates the X-ray spectra when BH LMXBs approach the Eddington limit. Throughout a single outburst, a LMXB usually experiences at least two of these states as the accretion rate onto the compact object evolves with time.

These transient LMXB outbursts are likely driven by thermal and viscous instabilities in the accretion disk \citep{dubus01, lasota01}. Both NS and BH LMXBs can be transient sources, although their outbursts are quantitatively different (see \citealp{done07} for a review). NS LMXBs generally host smaller accretion disks than their BH counterparts due to tidal truncation by the donor and the lower mass of the compact object. The smaller accretion disk is less likely to be unstable for a given donor star, and even when unstable, these smaller disks result in lower amplitude outbursts than those seen in BH LMXBs \citep{done07}. BH LMXBs can exhibit optical outbursts of $\gtrsim 5$~mags \citep{corral15}, and go years between consecutive outbursts \citep[e.g., ][]{russell18}. These outbursting LMXBs are well-described by the classical disk instability model (DIM) with modifications to account for self-irradiation \citep[DIM+irradiation, ][]{dubus01}.

\name was discovered by the All-Sky Automated Survey for SuperNovae (ASAS-SN, see \citealp{shappee14} and \citealp{kochanek17} for details on cameras, filters, and zero-points) on UT 2018-03-06.58 (MJD 58184.079861) at $\textrm{RA}=18^{\rm h}20^{\rm m}21.\!\!^{\rm{s}}9$  $\textrm{Dec.}=+07\degree11'07.\!\!''3$ (J2000) with a $V$-band magnitude of 14.88. It was publicly released within hours of discovery.\footnote{\url{http://www.astronomy.ohio-state.edu/asassn/transients.html}}  The source was undetected ($V > 16.7$ mag) on UT 2018-03-02.59, roughly 4 days prior. Because of its coincidence with a $G=17.8$ mag Gaia source \citep[ID 4477902563164690816, ][]{gaia1}, ASASSN-18ey was initially labeled as a cataclysmic variable (CV) candidate. Then, six days later on 2018-03-11, the Monitor of All-sky X-ray Image \citep[MAXI, ][]{MAXIref} Gas Slit Camera \citep[GSC, ][]{GSCref} nova alert system detected a bright X-ray transient at the same location (MAXI J1820+070, $32 \pm 9~\rm{mCrab}$, $4-10~\rm{keV}$, \citealp{ATel11399,ATel11400}). 

\begin{figure*}
    \centering
    \includegraphics[width=0.9\linewidth]{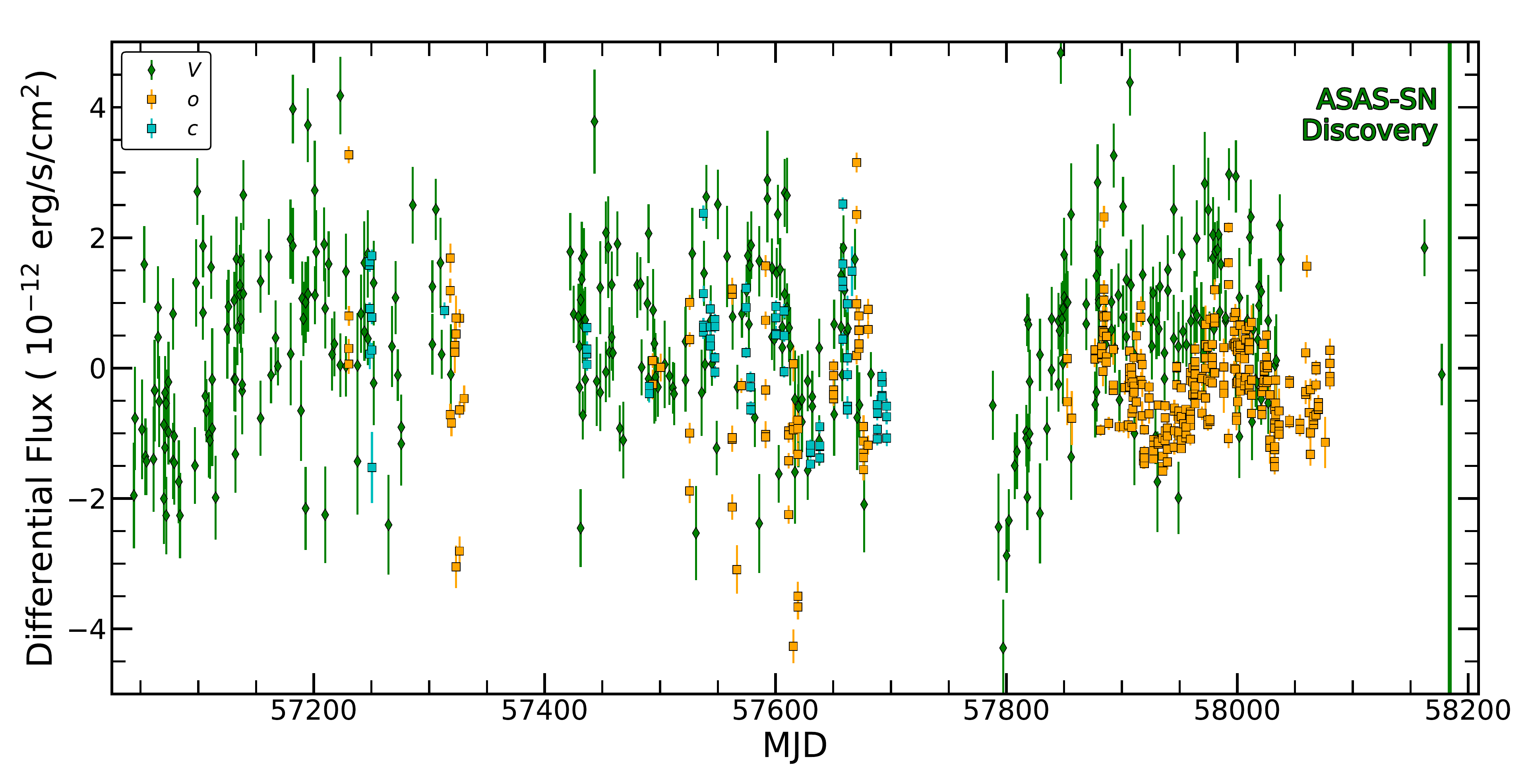}
    \caption{Pre-outburst light curve of \name from ASAS-SN and ATLAS. Green diamonds represent ASAS-SN $V$-band, blue squares ATLAS $c$-band, and orange squares ATLAS $o$-band observations. There are obvious correlations between observations conducted in different filters, indicating much of the variability is intrinsic to \name and not nightly scatter. Fluxes are \textit{not} corrected for interstellar extinction.}
    \label{fig:pre-outburst}
\end{figure*}

Several teams carried out follow-up observations after the MAXI transient alert due to the intrinsic brightness and unknown nature of \name. Follow-up X-ray observations of \name were conducted with NICER \citep{ATel11423, ATel11576, ATel11823}, INTEGRAL \citep{ATel11478, ATel11488, ATel11490}, and XRT/BAT \citep{ATel11403, ATel11427, ATel11578}. The first suggestion of \name being a BH LMXB and the detection of a possible state transition were posted by \citet{ATel11418} and \citet{ATel11820}, respectively. Optical spectra showed signatures typically associated with LMXBs in outburst including \ion{He}{1} and \ion{He}{2} in emission combined with an evolving H$\alpha$ emission profile \citep{ATel11424, ATel11425, ATel11480, ATel11481}. Subsequent optical observations revealed an optical period of $\sim 3.4~\rm{hr}$ \citep{ATel11596}, correlations between X-ray and optical brightness \citep{ATel11432, ATel11510, ATel11574}, linear polarization \citep{ATel11445}, and \newline(sub-)second flaring \citep{ATel11421, ATel11426, ATel11437,ATel11510, ATel11591, ATel11723, ATel11824}. These rapid photometric variations were confirmed in the near-infrared by \citet{ATel11451}, and a large IR excess compared to archival 2MASS data was noted by \citet{ATel11458} and \citet{ATel11855}. Radio observations may have detected a forming jet and its ensuing quenching \citep{ATel11420, ATel11439, ATel11440, ATel11539, ATel11540, ATel11609, ATel11827, ATel11831, ATel11887}.

Using the \textit{Gaia} DR2 parallax \citep{gaia1,gaia2,gaia3} and \citet{bailerjones18} distance priors, \name is located at a distance of $d = 3.06 ^{+1.54} _{-0.82}~\rm{kpc}$, with a total reddening of $E(B-V)=0.197$ mag \citep{SF11}, corresponding to $A_V = 0.614$ mag assuming $R_V=3.1$. We discuss pre-outburst data in \S\ref{sec:preoutburst}, analyze the rising light curves in \S\ref{sec:lc}, and qualitatively examine the pre- and post-maximum spectra in \S\ref{sec:spec_evol}. Finally, in \S\ref{sec:new_BHXB}, we discuss our conclusion that \name is likely a new BH LMXB in outburst.

\section{Quiescence }\label{sec:preoutburst}

Using archival photometry, we place limits on the maximum mass of the donor star in \name. We obtain $grizy$ magnitudes from the Pan-STARRS (PS) Stack Object Catalog\footnote{\url{http://archive.stsci.edu/panstarrs/stackobject/search.php}} \citep[SOC, ][]{chambers, flewelling16} and compare them to absolute magnitudes from the PARSEC\footnote{\url{http://stev.oapd.inaf.it/cgi-bin/cmd}} \citep{PARSECref} stellar isochrones. There are two entries in the PS SOC for \name, so we take the entry with the brighter $r$-band magnitude, although they are comparable in all filters. The source does not appear extended or blended in the PS catalog, and there are no other sources within several arc-seconds, so we consider the possibility of source confusion negligible. Using the $Gaia$ distance and PS magnitudes, we find a constraint on the donor star mass of $M_{\rm{donor}} \lesssim 1M_\odot$. Even  at the $3\sigma$ upper limit on the distance $d \simeq 7.7~\rm{kpc}$,  the donor star is still constrained to be a main sequence star with $M_{\rm{donor}} \lesssim 1.6 ~M_\odot$. As such, we can confidently rule out high-mass X-ray binaries (HMXBs, $M_{\rm{donor}} \gtrsim 10~M_\odot$) as potential candidates for \name, although this was already unlikely because HMXBs rarely experience outbursts \citep{done07}.

ASAS-SN and the Asteroid Terrestrial-impact Last Alert System (ATLAS, \citealp{tonry18}) observed the location of \name 403 and 707 times since January 2015 and September 2015, respectively.  ASAS-SN observes in $V$ and $g$ while ATLAS observes in orange ($o$) and cyan ($c$) filters. There is a bright ($V\sim 13.5$ mag) star roughly 18\farcs{0} from \name that contaminates both the ASAS-SN and ATLAS photometry in quiescence. During outburst, \name is bright enough that blending is not an issue. Thus, for the pre-outburst light curve we focus on changes in flux as shown in Figure \ref{fig:pre-outburst}.

\name has not had any comparable optical outbursts in the last $\sim 3.5$~years. The longest interval between any two consecutive ASAS-SN or ATLAS observations is 90 days, far shorter than the current outburst which has now surpassed 150 days. Assuming a conservative ASAS-SN detection limit of $5\times 10^{-12}$~erg/cm$^2$/s, we rule out outbursts with $\gtrsim 10\%$ of the amplitude of the current outburst since January 2015.  

The source shows noticeable variability in the pre-outburst light curve. We compute the fractional variability amplitude $f$ \citep{vaughn03} for each filter:

\begin{equation*}
    f_V = 0.51, \quad f_o = 0.27, \quad\rm{and}\quad f_c = 0.25 \quad (mag)
\end{equation*}

\noindent much of which we attribute to intrinsic variability in \name, especially given the correlations between filters. We inspected the periodograms for each light curve, but found no strong correlated peaks between them.  However, cross-correlating the light curves where observations in multiple filters are available shows the variability is correlated, with the correlation coefficient $r \sim 0.3-0.6$, depending on the MJD range and filter choices.

Photometric variability, even in quiescence, is expected for BH LMXB systems, and is sometimes attributed to a jet \citep[e.g. ][]{russell18}. Some BH LMXBs have shown flux increases in the optical and NIR on the order of $\sim 0.05~\rm{mag}~\rm{yr}^{-1}$ as they approach an outburst \citep[e.g. ][]{bernardini16, russell18}.  However, the ASAS-SN and ATLAS light curves are consistent with no systematic brightening since 2015. 

\begin{figure*}
\includegraphics[width=\linewidth]{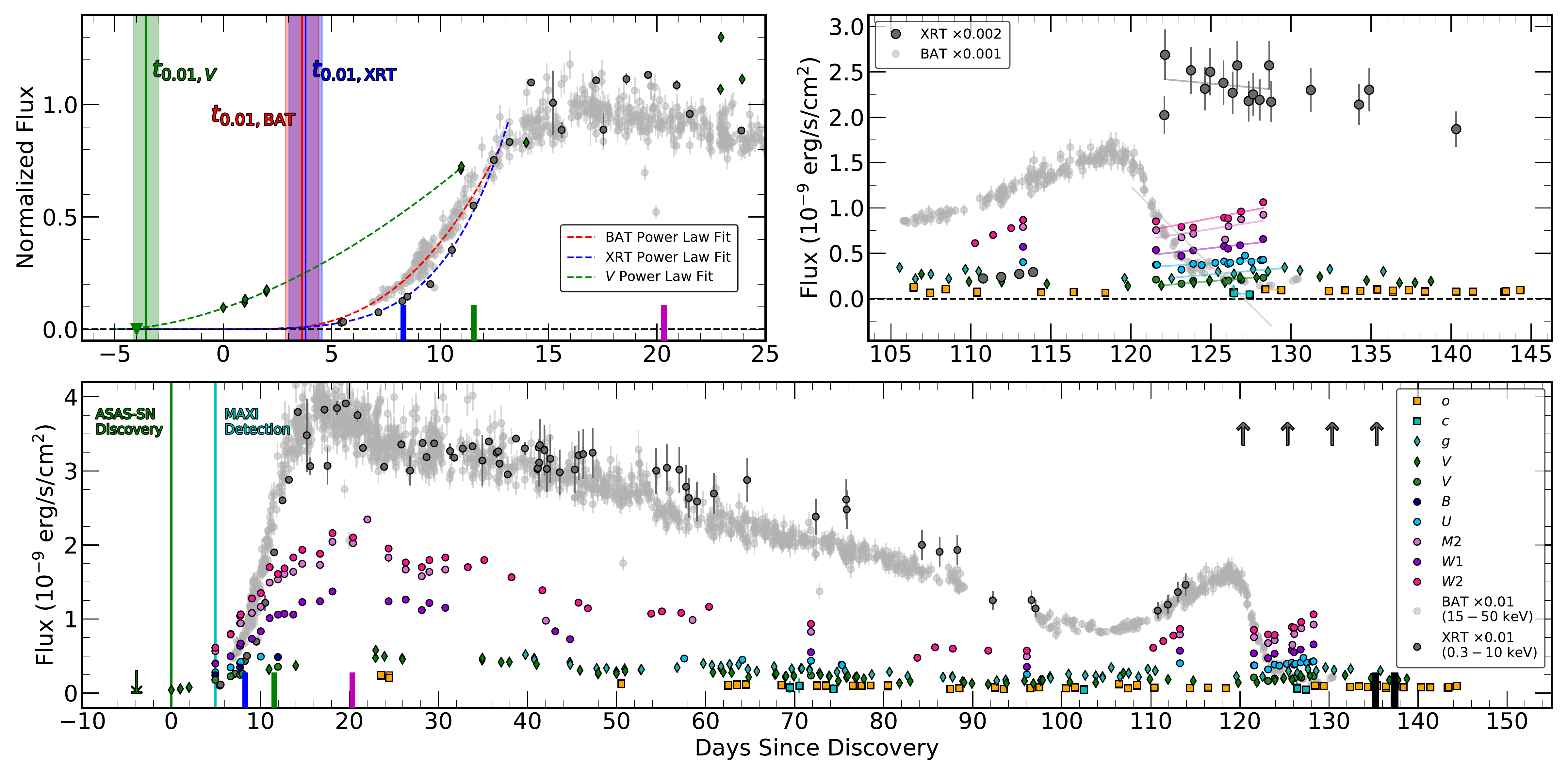}
\caption{Light curves of \name from discovery until $\sim 150~\rm days$ later in the optical, UV and X-rays. ASAS-SN photometry is plotted as diamonds, \swift as circles, and ATLAS as squares. XRT and BAT X-ray light curves are scaled for visual clarity.
\textit{Top Left}: Rising light curve of \name from ASAS-SN and $Swift$, including power-law fits and derived values of $t_{0.01}$ (see text). 
\textit{Top Right}: The hard X-ray re-brightening event at $\sim 120~\rm days$ after discovery. The optical and UV data imply a rising disk temperature (see Figure \ref{fig:LTRplot}) even though the hard X-rays have diminished and the soft X-rays are declining, indicating a delayed response between X-ray production and the subsequent re-processing into optical/UV emission (faint lines meant to guide the eye). The X-ray flux scales are \textit{not} the same as in the lower panel.
\textit{Bottom}: Full light curve of \name from ASAS-SN, ATLAS, and \swift. At this scale, the $Swift$ XRT fluxes at $>120$ days are off the top of the plot, as indicated by the upwards arrows. The green line indicates the discovery of \name and the cyan line marks the MAXI X-ray detection \citep{ATel11399}. Colored ticks along the x-axes indicate spectral epochs shown in Figure \ref{fig:spectra}.
}
\label{fig:lightcurve}
\end{figure*}

There is a $\sim 2.4\sigma$ X-ray detection by the ROentgen SATellite \citep[ROSAT, ][]{ROSATref} taken on MJD $48\,136$, almost 3 decades prior to the current outburst. The $373\,\rm s$ ROSAT observation has a count rate of $(4.42\pm3.16) \times 10^{-2}$ counts per second. Using the power law index and $n_H$ derived in \S\ref{sec:lc}, this corresponds to an un-absorbed flux of $(2.71\pm1.93)\times 10^{-12} ~\rm{erg}^{-1}\,\rm s^{-1}\,\rm{cm}^{-2}$ in the $0.3-2~\rm{keV}$ energy range ($L_{2-10\rm{keV}} \sim 10^{33}~\rm{erg}\,\rm s^{-1}$). If real, this implies an unusually high mass transfer rate from the donor, indicating the star has likely evolved off the main sequence. Archival photometry excludes most giant stars but subgiant stars are still possible considerations. There is also a 6 second slew-mode XMM-Newton observation of this location, but it provides weaker limits than the ROSAT observation.

\section{Outburst and Photometric Evolution}\label{sec:lc}

In Figure \ref{fig:lightcurve} we present ASAS-SN (filters: $V$, $g$), ATLAS (filters: $o$, $c$) and \textit{Swift Gamma-ray Burst Mission} \citep[\swift, ][]{gehrels04} UltraViolet and Optical Telescope \citep[UVOT, filters: $v$, $b$, $u$, $W1$, $M2$, $W2$;][]{roming05}, X-ray Telescope \citep[XRT, $0.3-10~\rm{keV}$ ; ][]{burrows05} and the Burst Alert Telescope \citep[BAT, $15-50~\rm{keV}$][]{BATref} observations covering
 from discovery until $\sim 150~\rm days$ after discovery (\swift PIs: Kennea, Motta, Paice, Tanaka, Altamirano, Sanchez, Yan, Markwardt, Yu, Sivakoff, Knigge). As each UVOT epoch contained 2 observations in each filter, we first combined the two images in each filter using the HEAsoft software task {\tt uvotimsum}, and then extracted counts from the combined images in a 10\farcs{0} radius region using the software task {\tt uvotsource}, with a sky region of $\sim$~40\farcs{0} radius used to estimate and subtract the sky background. Aperture corrections were applied to the UVOT count rates before converting into magnitudes and fluxes based on the most recent UVOT calibration \citep{poole08,breeveld10}. The brightness of \name caused several \swift $b$-band observations to be saturated, as well some saturated images in other bands, which we exclude for the entirety of our analysis. The BAT data was retrieved through the BAT Transient Monitor\footnote{\url{https://swift.gsfc.nasa.gov/results/transients/}} \citep{BATmonitor} and the XRT data was retrieved through the UK \swift Science Data Centre\footnote{\url{http://www.swift.ac.uk/user_objects/}} \citep{evans07, evans09}. These tools automatically handle processing steps such as source extraction and background subtraction, and mitigate potential issues such as detector pile-up for bright sources.
 
We fit the 54 XRT spectra simultaneously using XSPEC, assuming an absorbed power-law model to derive the column density along the line-of-sight, $N_H = 1.05 \times 10^{21}~\rm{cm}^{-2}$, and a photon index of $\Gamma = 1.4$, which we use in converting count rates to fluxes with WebPIMMS\footnote{\url{https://heasarc.gsfc.nasa.gov/cgi-bin/Tools/w3pimms/w3pimms.pl}}. This methodology neglects the changing spectral forms of \name as it undergoes the outburst, however, for our purposes, this is a sufficient treatment. This column density implies a reddening of $E(B-V) \simeq 0.18$, consistent with the estimate reported in \S\ref{sec:intro} from \citet{SF11}. All photometry was converted to flux units (erg/cm$^2$/s) for comparison with the X-ray data and corrected for interstellar extinction/absorption. Quiescent optical fluxes, derived from the PS SOC magnitudes using conversions in \citet{tonry12} and \citet{tonry18}, were added to the ASAS-SN and ATLAS differential flux measurements to more accurately represent the true optical flux from \name. 

\subsection{Rising Light Curve}

The optical rises of BH LMXBs prior to the corresponding X-ray detection are generally poorly constrained, although there are a few occurrences \citep[e.g., ][]{orosz97, jain01, zurita06}. Similarly, ASASSN-18ey was discovered in the optical, allowing us to study its pre-X-ray transient evolution. To characterize the rising light curves we use a power law, 

\begin{equation}\label{eq:risingLC}
    \hat{F}_\lambda(t) = A\bigg(\frac{t-t_{0,\lambda}}{\textrm{1 day}}\bigg)^B 
\end{equation}

\noindent where $\hat{F}_\lambda$ is the normalized flux in filter $\lambda$, $t_{0,\lambda}$ is the time zero point, $t$ is the time of the observations, and $\{A, B\}$ are coefficients. For the rising $V$-band light curve, only the ASAS-SN observations are used, as the \swift $v$-band is both bluer and narrower than the Johnson-Cousins $V$ filters used by ASAS-SN.  Since the functional form we adopt is not physically motivated and there is quiescent flux in the optical, we report the time $t_{0.01}$ at which the light curve for each filter/energy range reaches 1\% of the peak flux rather than $t_{0}$.  This time is closer to the data being fit and is therefore less sensitive to the exact functional form we assume. 

The power-law fits to the rising ASAS-SN $V$-band, $Swift$ XRT, and $Swift$ BAT light curves are shown in the top left panel of Figure \ref{fig:lightcurve}, along with the derived values of $t_{0.01}$. We use a bootstrap-resampling technique to estimate errors for the fit parameters since the sampling of the rising light curve is sparse. The ASAS-SN $V$-band observation at $\sim 14~\rm{days}$ after discovery is excluded from the fitting because it significantly reduces the quality of the fit without affecting the estimate of $t_{0.01,V}$ appreciatively. For the ASAS-SN $V$-band light curve, we find $A = (5.85\pm2.07)\times 10^{-3}$, $B=1.74\pm0.11$, and $t_{0.01,V} = -3.56\pm0.57~\rm days$. For the XRT light curve, we find $A = (5.79\pm7.18)\times 10^{-8}$, $B=5.82\pm0.35$, and $t_{0.01,\rm{XRT}} = 3.80\pm0.77~\rm days$. For the BAT light curve, we find $A = (1.42\pm1.10)\times 10^{-3}$, $B = 2.62\pm0.25$, and $t_{0.01,\rm{BAT}} = 3.64\pm0.78~\rm days$. Thus, we find a lag between the start of the optical and BAT hard X-ray outbursts of $7.20 \pm 0.97$~days. 

We fit quadratic functions to determine the time of maximum, $t_{\rm{max}}$, in days relative to discovery for the ASAS-SN $V$-band, XRT and BAT light curves. We then use the derived $t_{\rm{max}}$ to calculate the flux at peak, which we use in normalizing the rising light curves. We find $t_{\rm{max}}=16.86\pm0.73$~d for BAT, $18.06\pm3.99$~d for XRT, and $26.20\pm1.52$~d for ASAS-SN $V$-band. Combining these with our derived $t_{0.01}$ values, we find total rise times $t_{\rm{rise}} = t_{\rm{max}} - t_{\rm{0.01}}$ of $13.22\pm 1.07$~d, $14.26\pm 4.07$~d, and $29.76\pm1.63~\rm days$ for the BAT, XRT, and ASAS-SN $V$-band, respectively. While the optical rise begins first, the optical luminosity takes $16.54\pm 1.95$~days longer to peak than the hard X-rays. 

This week-long delay between the optical and X-ray flux increases is expected from disk instability models and can constrain where in the disk the outburst began \citep{dubus01}. The outburst starts in the $V$-band at radius $R(V)$ and simultaneously propagates both inwards and outwards until reaching the inner regions of the disk, producing X-rays. The difference between the beginning of the optical and X-ray emission corresponds to the viscous timescale $t_{\rm{visc}} = \Delta t_{V -X} = t_{0.01,V} - t_{0.01,\rm{BAT}}$, and constrains the location in the disk at which the outburst begins. We adopt the scaling relation of \citet{bernardini16}, using the same parameter ranges for the hot disk viscosity parameter ($\alpha \in [0.1-0.2]$), mid-plane disk temperature ($T_s \in [3-5]\times 10^4~\rm K$), and assumed radius of the X-ray disk ($R(X) = 5\times 10^8~\rm{cm}$). The BH mass is currently unconstrained, but the dependence is weak ($t_{\rm{vis}}\propto \sqrt{M_{\rm{BH}}/10 M_\odot}$), so we assume $M_{\rm{BH}} \approx 8M_\odot$ for simplicity. Taking $t_{\rm{vis}} \in [6.1-8.1]~\rm days$ ($t_{\rm{vis}} \pm 1\sigma$), we find $R(V) = [0.8-2.5]\times 10^9~\rm{cm} = [0.01-0.04]~R_\odot$. This is consistent with the range found for V404 Cyg by \citet{bernardini16} and matches theoretical predictions for DIM+irradiation models for BH LMXB outbursts \citep{dubus01}. 

\subsection{Temperature and Photosphere Evolution}

\begin{figure}
    \centering
    \includegraphics[width=\linewidth]{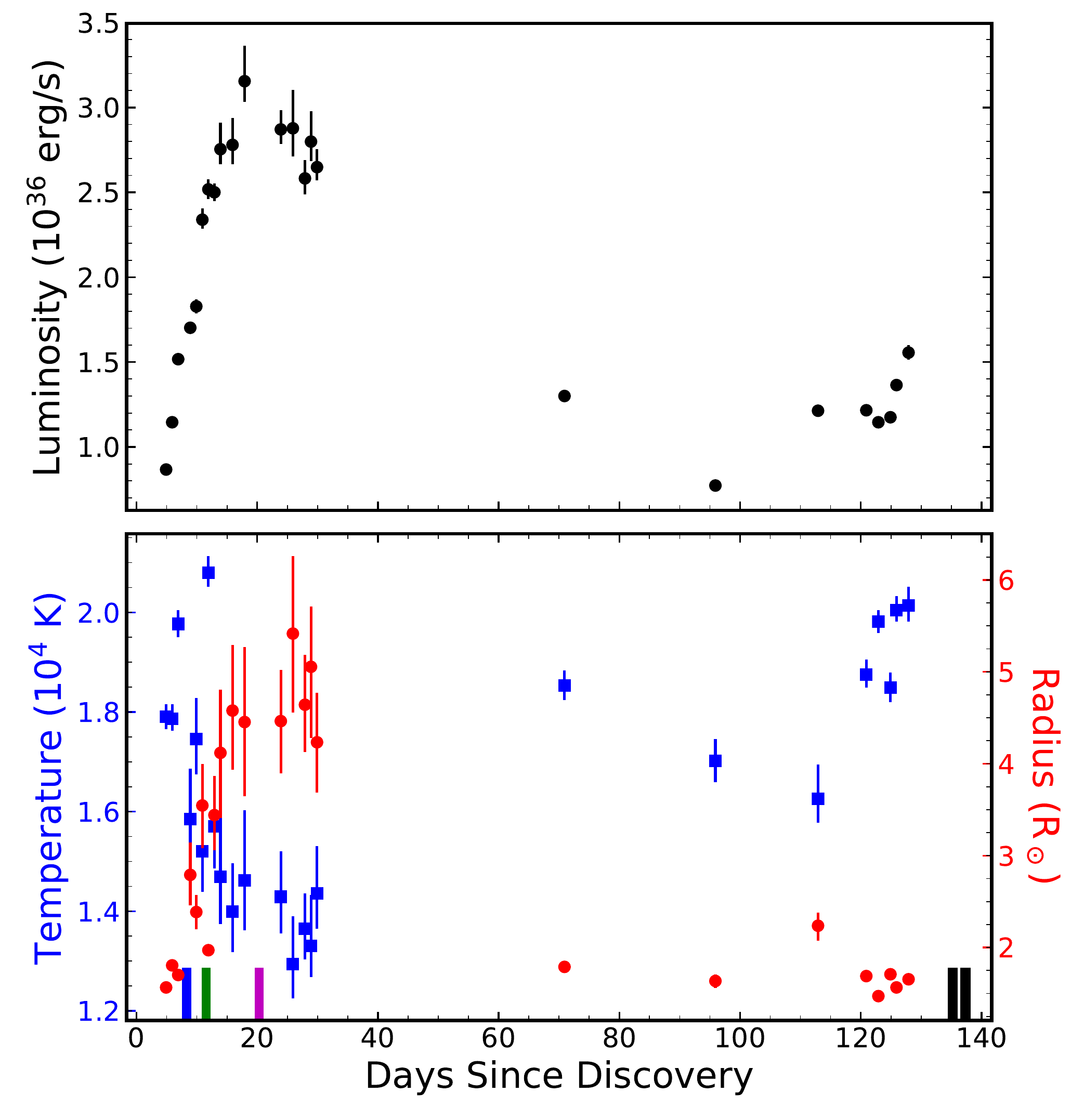}
    \caption{Luminosity (top), radius, and temperature (bottom) evolution of \name from blackbody fits to the $Swift$ photometry. Spectral epochs are shown along the bottom axis, color-coded according to the spectra in Figure \ref{fig:spectra}.}
    \label{fig:LTRplot}
\end{figure}

\begin{figure*}
\includegraphics[width=\linewidth]{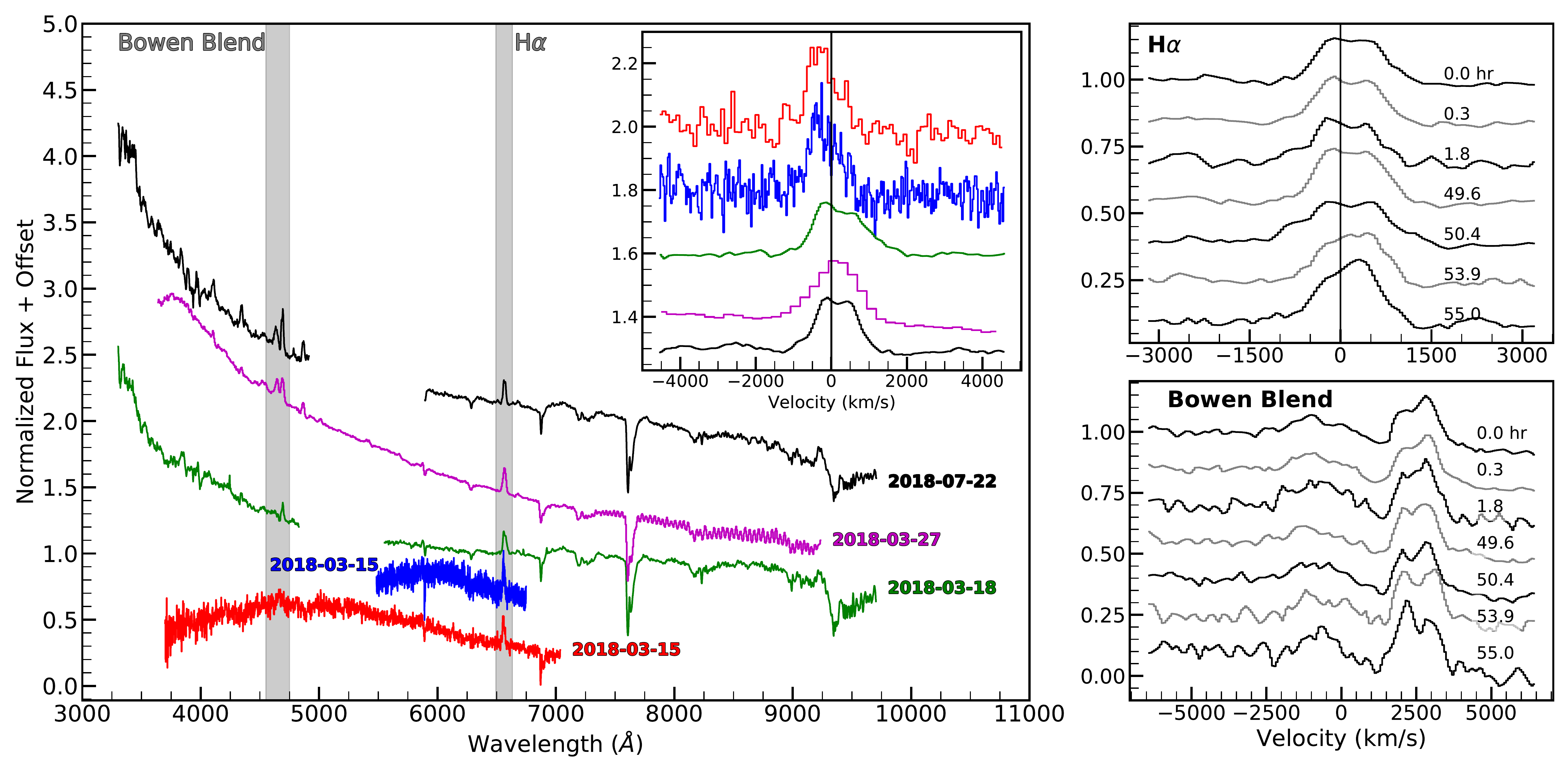}
\caption{\label{fig:spectra} Spectroscopic evolution of \name.  \textit{Left}: Pre-maximum spectra of \name (colored) and at $135~\rm days$ (black). The colors correspond to those used for the colored ticks along bottom axes of Figures \ref{fig:lightcurve} and \ref{fig:LTRplot}. \textit{Left Inset}: The region around H$\alpha$ showing the broad, asymmetrical emission line that presents the first evidence of a double-peaked profile in the UH88 spectrum taken on 2018-03-18. \textit{Right Panels}: Seven late-time ($ 135-137~\rm days$) spectra of \name zoomed into the region around H$\alpha$ (top) and the Bowen blend (bottom) showing the evolving double-peaked profiles. Numbers correspond to hours after the first late-time spectrum. 
}
\end{figure*}

We compute the effective blackbody (BB) luminosity, temperature and radius as a function of time for epochs with at least 3 $Swift$ filters. An accretion disk is not usually well modeled by a simple BB, the spectral energy distribution in the observed photometry is well fit by a blackbody model, suggesting that a narrow range of temperatures dominate the peak of the emission. These BB parameters are presented in Figure \ref{fig:LTRplot} and errors derived using MCMC.  The overall temperature of the disk is consistent with the ionization temperature of hydrogen, as expected for an outburst driven by thermal and viscous instability \citep{dubus01}. The BB evolution show the photosphere to be cooling and expanding as \name approaches its peak optical luminosity, consistent with an outward-propagating heating front moving through the disk.

Unfortunately, the \swift data has a gap from $\sim 35-70~\rm days$ after discovery where we cannot constrain the BB evolution.  During this gap the effective temperature of the re-processing disk increases and the photosphere recedes but we cannot constrain how quickly this change occurred.

We now have a rough picture of the structure and evolution of \name as it undergoes the outburst. The initially hot and small emission region cools and expands as \name reaches peak optical brightness $\sim 30~\rm days$ after discovery. Post peak, the photosphere shrinks and becomes hotter before starting another cooling phase. It is worth noting that this secondary cooling phase is dissimilar to the original, as the size of the emission region remains small. At the end of the linear X-ray decay, when \name has settled into a quasi-static ``plateau'', the photosphere has reverted to a temperature and size similar to those before the peak. By $\sim 120~\rm days$ after peak, \name has started rising in temperature, consistent with the rise in UV brightness, even as the hard X-rays turn off and the soft X-rays diminish, indicating a delayed response between X-ray production and the response of the re-processing disk.

\section{Spectral Evolution}\label{sec:spec_evol}

In Figure \ref{fig:spectra} we show the spectroscopic evolution of \name from 8 to 137 days after discovery. We present two early-time SOAR/Goodman spectra \citep{clemens04} acquired roughly 8 days after discovery on MJD 58192.4 taken with two separate gratings: a $600~\rm s$ low-resolution spectrum spanning $3700-7000$\AA{} and a $1200~\rm s$ high-resolution spectrum spanning $5500-6800$\AA.  These spectra were reduced using standard \IRAF{} routines.  We also obtained one early time (MJD 58195.6) and seven late-time (MJD 58319.3-58321.6) spectra with the University of Hawaii 88'' (UH88) telescope and the SuperNova Integral Field Spectrograph \citep[SNIFS, ][]{SNIFSref}. Each spectrum was reduced with an automated pipeline and spans $3300-9700$\AA{} excluding the dichroic crossover ($\sim 4800-5500$\AA). Finally, we supplement our spectroscopic time series with a publicly available ePESSTO \citep{ATel11480, smartt15} spectrum taken near peak.

\begin{figure*}
    \centering
    \includegraphics[width=0.9\linewidth]{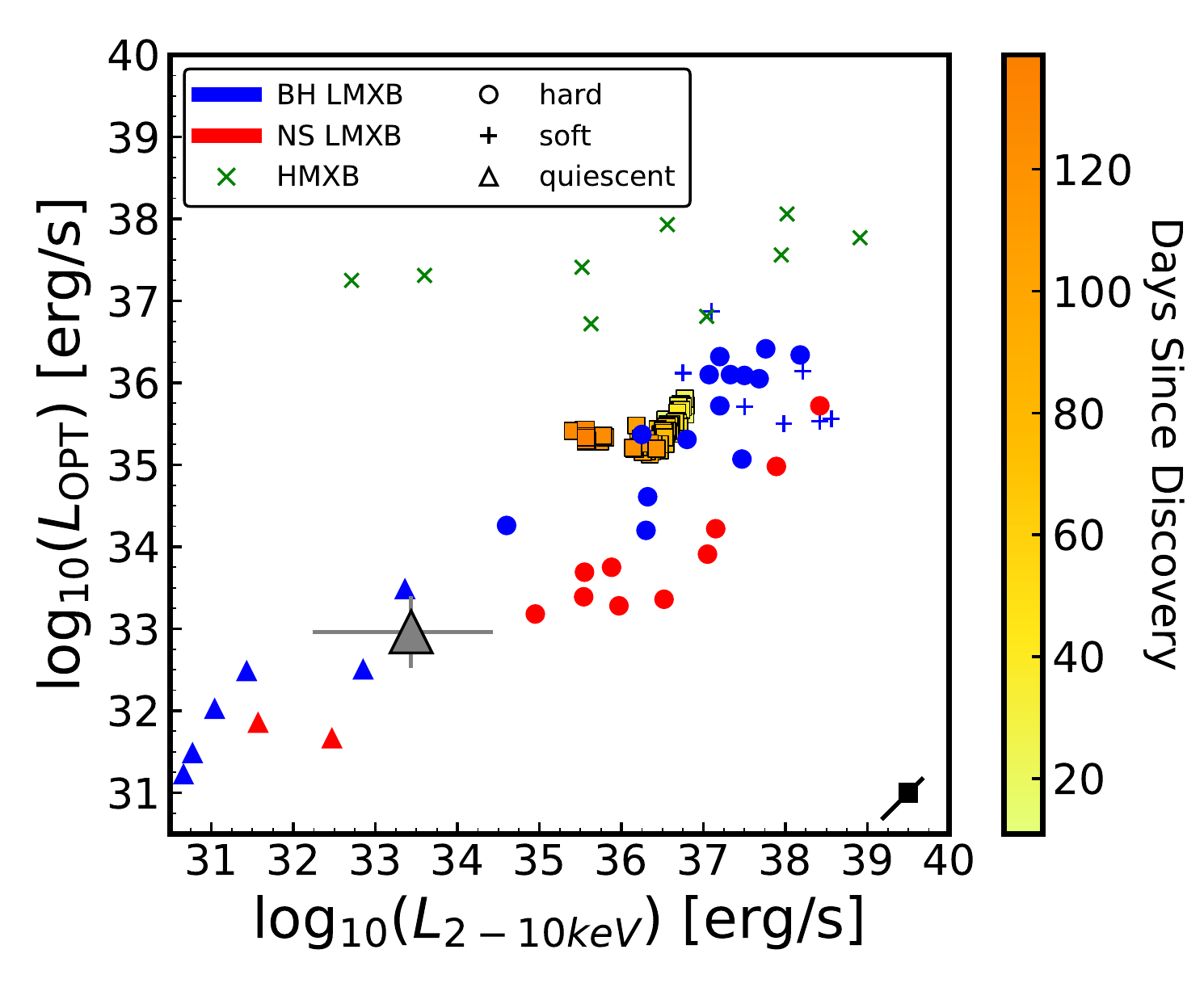}
    \caption{Evolution of \name in the X-ray vs. optical luminosity diagram along with various categories of X-ray binaries taken from \citet{russell06} for comparison. \name coincides with the region occupied by BH LMXBs throughout the outburst. The gray triangle indicates the rough position of \name assuming the pre-outburst ROSAT X-ray detection and optical brightness derived from PS $g$-band photometry. The black square indicates the correlated uncertainty due to the distance. The flux measurement uncertainties during outburst are negligible on this scale, and see \S\ref{sec:new_BHXB} for an explanation of the quiescent uncertainties.}
    \label{fig:LxLopt}
\end{figure*}

The spectra exhibit largely featureless continua except for the Balmer lines and the Bowen blend, a forest of \ion{C}{3} and \ion{N}{3} lines between $\sim 4630-4660$~\AA{} created by high-energy irradiation of the companion star \citep{mcclintock75}. A striking feature of the spectroscopic evolution is the development of a double-peaked H$\alpha$ profile (inset, Figure \ref{fig:spectra}) between 8 and 11 days after discovery.  Another noteworthy feature in the SNIFS pre-maximum spectrum is the appearance of a single-peaked emission line in the Bowen blend region, which becomes double-peaked $\sim 9~\rm days$ later.

Finally, we highlight the short-time-scale variations of H$\alpha$ and the Bowen blend in the late time spectra of \name (right panels, Figure \ref{fig:spectra}). The seven spectra, taken over 2.5 days spanning 135-137 days after discovery, exhibit shifting line profiles. Even consecutive exposures separated by $< 30$~minutes show qualitative shifts in the H$\alpha$ and Bowen blend, likely corresponding to the orbital motion of the system.

\section{A New Black-Hole X-ray Binary}\label{sec:new_BHXB}

Figure \ref{fig:LxLopt} shows the location of \name on the $L_X-L_{\rm{opt}}$ diagram from \citet{russell06}. Due to the high cadence optical and X-ray observations throughout the outburst, we can track the temporal evolution of \name. Throughout the outburst, \name resides in a region occupied by BH LMXBs in the hard state. Although the pre-outburst ROSAT detection has low significance, the inferred X-ray luminosity and the optical luminosity agree with BH LMXBs in quiescence (Figure \ref{fig:LxLopt}, gray triangle). The uncertainties on the quiescent X-ray flux are propagated from the low significance detection, with an additional contribution from different assumptions about the photon index of BH LMXBs in quiescence from \citet{remillard06}. The optical uncertainties stem from choosing different filters from the archival PS SOC photometry to calculate the quiescent luminosity (see \S\ref{sec:preoutburst}).

\name is almost certainly a new BH LMXB. The pre-outburst optical light curve is intrinsically variable, consistent with an unstable accretion system. The quiescent X-ray and optical fluxes preclude HMXBs, persistent NS LMXBs, and all giant companions. An outburst of $\Delta V\sim 6$~mags matches known BH LMXB outbursts \citep{corral15} and precludes most non-BH systems where the accretion disks are less likely to be unstable and thus less likely to experience an outburst \citep{done07}. The peak of the soft and hard X-ray light curves precedes the UV, which in turn precede the optical peaks, providing further evidence of an accretion disk that is re-processing the accretion-generated X-rays and increasing in temperature. \citet{veladina18} conducted polarimetric observations of \name in outburst, finding only a minimal amount of polarized flux ($< 1\%$), consistent with the majority of optical flux stemming from the disk. The temperature of the disk ($\sim 10^4~\rm K$) and $\sim7~\rm day$ delay between the optical and X-ray rise are consistent with an outburst driven by H ionization instability.  The spectra show typical characteristics of LMXBs in outburst, such as variable H$\alpha$ and Bowen blend profiles. The evolution of \name on the $L_X-L_{\rm{opt}}$ diagram during the outburst further strengthens the classification as a new black-hole X-ray binary, as it resides in a region dominated by BH LMXBs for the entirety of the outburst. 

Once the system has returned to quiescence and the mass of the BH is determined, we can use $t_{\rm{vis}}$ to constrain, rather than assume, the temperature and viscosity of the hot disk at the beginning of the outburst. This demonstrates the benefit of catching BH LMXBs outbursts on the rise. \name has the potential to be the best-studied BH LMXB outburst to-date, with more than 40 Astronomer's Telegrams and $> 360\,000$ observations from 50 observers reported on the AAVSO Light Curve Generator\footnote{\url{https://www.aavso.org/lcg}} as of 2018 October 1 \citep{AAVSOref}.

\vspace{1cm}
\section*{Acknowledgements}

We thank the referee for constructive comments which improved this manuscript. We also thank Connor Auge, Gagandeep Anand, and Aaron Do for useful discussions. MAT acknowledges support from the United States Department of Energy through the Computational Sciences Graduate Fellowship (DOE CSGF). DMR acknowledges support from Research Experience for Undergraduate program at the Institute for Astronomy, University of Hawaii-Manoa funded through NSF grant AST-1560413. CSK and KZS are supported by NSF grants AST-1515876 and AST-1515927. SD acknowledges Project 11573003 supported by NSFC. Support for JLP is provided in part by the Ministry of Economy, Development, and Tourism's Millennium Science Initiative through grant IC120009, awarded to The Millennium Institute of Astrophysics, MAS. TAT is supported in part by Scialog Scholar grant 24215 from the Research Corporation. JFB is supported by NSF grant PHY-1714479.

We thank the Las Cumbres Observatory and its staff for its continuing support of the ASAS-SN project. ASAS-SN is supported by the Gordon and Betty Moore Foundation through grant GBMF5490 to the Ohio State University and NSF grant AST-1515927. Development of ASAS-SN has been supported by NSF grant AST-0908816, the Mt. Cuba Astronomical Foundation, the Center for Cosmology and AstroParticle Physics at the Ohio State University, the Chinese Academy of Sciences South America Center for Astronomy (CASSACA), the Villum Foundation, and George Skestos.

This work has made use of data from the European Space Agency (ESA) mission {\it Gaia} (\url{https://www.cosmos.esa.int/gaia}), processed by the {\it Gaia}
Data Processing and Analysis Consortium (DPAC,
\url{https://www.cosmos.esa.int/web/gaia/dpac/consortium}). Funding for the DPAC
has been provided by national institutions, in particular the institutions
participating in the {\it Gaia} Multilateral Agreement.

Based on observations obtained at the Southern Astrophysical Research (SOAR) telescope, which is a joint project of the Minist\'{e}rio da Ci\^{e}ncia, Tecnologia, Inova\c{c}\~{o}es e Comunica\c{c}\~{o}es (MCTIC) do Brasil, the U.S. National Optical Astronomy Observatory (NOAO), the University of North Carolina at Chapel Hill (UNC), and Michigan State University (MSU).

This work made use of data supplied by the UK Swift Science Data Centre at the University of Leicester.




\bibliographystyle{mnras}
\bibliography{ref} 

\label{lastpage}
\end{document}